# Investigating the impact of copper-PEDOT:PSS matrix towards non-enzymatic electrochemical creatinine detection

Chirantan Das[1,4]*, Subhrajit Sikdar[2], Shreyas K. Vasantham[1], Piotr Pięta[1], Marcin Strawski[3], Marcin S. Filipiak[4], Paweł Borowicz[1], Yurii Promovych[1], Piotr Garstecki[1]*

[1]Institute of Physical Chemistry, Polish Academy of Sciences, 01-224 Warsaw, Poland

[2]Department of Electronic and Electrical Engineering, University of Sheffield, Sheffield, S1 3JD, United Kingdom

[3]University of Warsaw, Faculty of Chemistry, Pasteura 1, 02-093 Warsaw, Poland

[4]Centre for Advanced Materials and Technologies CEZAMAT, Warsaw University of Technology, Poleczki 19, 02–822 Warsaw, Poland

*Corresponding authors.

Email address: chirantan.das@pw.edu.pl (C. Das), pgarstecki@ichf.edu.pl (P. Garstecki).

## Abstract

Electrochemical creatinine sensors offer great promise towards rapid, reagent-free and point-of-care (POC) kidney-function monitoring. However, challenges related to analyte binding, data reproducibility, sensitivity, fouling and device degradation deter its widespread implementation. Here, we show how a carbon electrode modified with a combination of poly(3,4-ethylene dioxythiophene): poly(styrene sulfonate) (PEDOT:PSS) and copper nanoparticles (CuNPs) can rapidly and accurately detect creatinine (CT) in artificial urine media employing electrochemical techniques. Applying redox potential sweeps (vs Ag/AgCl) using copper sulfate ($CuSO_4$) solution on such sensor facilitates the CuNP embedding process inside the conjugated polymer matrix which has been further validated by supporting techniques. We predicted and validated the formation and contribution of two Cu-CT coordination complexes corresponding to Cu(I) and Cu(II) states, which are responsible for CT detection. The fabricated CT sensor exhibits high selectivity against major artificial urine interferents and is stable for a month showing minimal degradation (0.53%) in peak current value. Such sensors can be utilized to detect and monitor different stages of renal failure in real-time patient samples.

# 1. Introduction

Chronic kidney disease (CKD) is characterized by a persistent alteration in the kidney structure and function, or both, with serious implications on the health of an individual[1-4]. Individuals affected by kidney diseases encompassing acute kidney injury (AKI), CKD and post-renal treatment kidney failure exceed 850 million globally[5-7]. CKD has undeniably emerged as a critical global public priority which is spreading at an alarming rate and therefore, require early diagnosis and intervention to control. From a clinical perspective, the classification and risk stratification of CKD heavily hinge on the estimation of glomerular filtration rate (GFR) and the evaluation of urine albumin-to-CT ratio (uACR)[8,9]. CT (2-amino-1-methyl-5H-imidazol-4-one) is the final product of creatine phosphate metabolism occurring in muscles which is responsible for providing energy to muscle tissues and is the most important indicator of different stages of kidney diseases[10,11]. Quantitative determination of CT in blood serum [12,13], urine [13,14] and in recent times, sweat [15], has been a significant clinical measurement factor for the diagnosis of renal dysfunction, muscle damage and thyroid malfunction [16]. For a healthy individual, the normal GFR ranges 90-100 mL/min/1.73m$^2$ and, the overall blood serum (sCr) and urine CT concentrations are within the range of 60-110 µmol/L and 7-16 mmol/L[8]. Abnormalities reflected in such ranges depict the onset of degradation in the bodily waste product filtration process through kidneys. Therefore, quantification of CT levels is a significant issue for clinical diagnosis which directly reflects on the quality of kidney function.

Historically quantification of CT began in 1886 when Max Jaffe discovered the reaction of CT with picric acid in an alkaline environment. This discovery laid the foundation for the Jaffe method, which, clinicians and researchers have relied on due to its simplicity and cost-effective CT estimation[11,16-17]. Apart from that, other conventional methods for CT detection involve high-performance liquid chromatography-mass spectrometry (HPLC-MS)[18,19], liquid chromatography-tandem mass spectrometry (LC-MS/MS)[20,21] and electrochemical enzyme-based sensing[22-24]. HPLC separates CT from other metabolites before spectrometric analysis, while LC-MS/MS uses stable isotope dilution for CT measurement.

In Jaffe methods, CT forms a yellow-orange complex after reacting with picric acid in alkaline conditions. The dye concentration is a proportional indication of the CT concentration. However, this technique suffers from major limitations due to the interference of other substances in the sample, lack of specificity, and its reagent intensive driving further costs and complexity[25]. Furthermore, the presence of co-morbidities such as diabetes and the use of antibiotics or other medications apart from renal failure influence the measured sCr levels causing delays in therapeutic solutions. In addition, blood collection necessitates a skilled person within a clinical facility or hospital, and the analysis can take several hours to days, impeding diagnosis, especially when the kidney function significantly deteriorates within a short span. Also, in cases of aged patients, clinical complications arise due to blood drawing which includes difficulty finding veins, diminished blood flow due to dehydration, fragile veins that are prone to injury and reduced blood volume[26]. Further, HPLC and LC-MS/MS methods are often time-consuming and require removal of interfering molecules, such as using sodium dodecyl sulfate to eliminate bilirubin interference [10,11]. Thus, there is an inherent need to develop cost-effective, portable, rapid, and specific methods to quantify CT, non-invasively.

In recent times, enzymatic biosensors offer excellent specificity and selectivity incorporating enzymes such as CT deiminase (CD), creatine amidohydrolase (CA) and sarcosine oxidase (SO) to catalyse the reaction for CT detection, but also suffer from constraints such as the complexity in immobilization onto the electrode surface, pH dependent stability, storage, and high costs[10,11,24,27-33]. In this context, non-enzymatic sensors offer several advantages such as lower fabrication costs, robust operation conditions, high selectivity and specificity owing to the target material-specific modification of the electrode and understanding the specific complex formations occurring at the electrode-solution interface[34-48]. Recently, several works towards non-invasive detection of CT have been reported and their key features have been summarized[34-48] in **Table 1** for comprehensive state-of-the-art comparison. Such reports show targeted employment of hybrid materials, developed from polymer-associated metals or metallic oxides[34,35,37,41], metal imprinted polymers (MIP)[37,38,42,44], multi walled carbon nanotubes (MWCNT)[39,40,46], metal nanoparticles (Au, Ni, Cu, Ag)[35,40,41,45] and quantum dots (CdS)[47], in various forms including thin films, micro/nanofibers, and wires which has been found effective in addressing challenges faced by conventional chemical sensors. However, electrochemical potential sweep modulated metal particle embedding inside conjugated polymer matrix for electrode modification has recently emerged to be a rapid process of physisorption of targeted materials[49], which can be a potential method in fabricating electrochemical CT sensors for CKD monitoring.

In this study, we present an electrochemical surface modification process on commercially available carbon-based screen-printed electrode (SPE) for the detection of CT in artificial urine media. Initially, a carbon working electrode is modified with PEDOT:PSS conductive polymer to realize the idea of output current amplitude enhancement. Such modification shows a significant two-order increase in output current value ($10^{-4}$ A) compared to the reference undoped carbon ($10^{-6}$ A). This is followed by electrochemical doping of such a polymer matrix with CuNPs (Cu@PEDOT:PSS) using potential sweeps. Post Cu embedding process, we observed metallic Cu to be physically adsorbed in the interstitial sites or pores of the conjugated polymer matrix. We further investigated the embedding process of Cu by employing typical characterization tools. The electrochemical behaviour of Cu@PEDOT:PSS-CT complex is also studied in detail using cyclic voltammetry (CV) and differential pulse voltammetry (DPV) techniques. In such voltammograms two opposite trends have been observed with increasing CT concentration which is correlated to the formation of two Cu@PEDOT:PSS-CT complexes. Comprehensive understanding of such complexes could be a potential factor for non-invasive CT detection which might be utilized to monitor different stages of renal abnormalities, thereby, facilitating clinical diagnostic processes.

**Table 1:** Summary of methods used, transducer material, nanomaterials for electrode modification, linear range of detection, Limit of Detection, Response Time of the measurements and samples considered for non-enzymatic electrochemical creatinine sensing.

| Methods used | Transducer | Nanomaterial | Linear Range (μM) | Sensitivity (μA.mM cm$^{-2}$) | LOD (μM) | Response time | Sample | References |
|---|---|---|---|---|---|---|---|---|
| CV/DPV | PAA gel–Cu(II)/Cu$_2$O NPs/SPCE | Cu$_2$O NPs | 200–100 000 | - | 6.5 | — | Urine | 34 |
| SWV | rGO/Ag NPs/GCE | Ag NPs, rGO | 1–12 | 0.04 | 74.3 | — | Urine | 35 |
| Amperometry | CNT–ABTS/Nafion®/SPCE | CNTs | 0–21 300 | 27.3 | 11 | 50 s | Urine | 36 |
| Amperometry | CuO@MIP/CPE | CuO NPs | 0.5–200 | 0.265 | 0.083 | — | Urine | 37 |
| EIS | MIP/carbon paste electrode | — | 0.18–5.92 | - | 0.18 | — | Human serum and artificial urine | 38 |
| DPV | CP/MWCNT–Inu–TiO$_2$ | TiO$_2$ NPs, MWCNTs | 0.2–1 | 0.2 | 0.06 | — | Human serum | 39 |
| | | | 50–12 000 | | 90 | 20 s | Urine | |
| DPV | Ag NPs/MWCNTs/FA/CPE | Ag NPs, MWCNTs | 0.01–200 | - | 0.008 | 1.5 s | Human serum and urine | 40 |

| Method | Electrode | Nanomaterial | Linear range | Sensitivity | LOD | Response time | Sample | Ref |
|---|---|---|---|---|---|---|---|---|
| CV | Cu NPs/PDA–rGO–NB/GCE | Cu NPs, rGO | 0.01–100 | - | 0.002 | — | Human serum and urine | 41 |
| DPV/EIS | MIP/Au–SCPE | — | 0.00088–0.00884 | - | 0.00014 | — | Urine | 42 |
| CV | Phosphotungstic acid/poly(ethylene imine)/ITO | C | 0.125–62.5 | - | 0.06 | 20 s | Urine | 43 |
| DPV | MIP/Ni@PANI NPs/MCGE | Ni@PANI NPs | 0.004–0.8 | - | 0.0002 | — | Urine | 44 |
| DPV | $Fe^{3+}$/CB NPs/SPCE | CB NPs | 100–6500 | - | 43 | — | Urine | 45 |
| DPV | Pectin–MWCNT/CPE | MWCNTs | 0.016–3.3 | 3.5 | 0.6241 | — | Urine | 46 |
| DPV | CdS quantum dots/PGE | CdS quantum dots | 0.442–8840 | - | 0.229 | — | Human serum and urine | 47 |
| DPV | $CDs/WO_3$@GO/GCE | CDs | 0.0002–0.112 | - | 0.0002 | — | Human blood and urine | 48 |
| **DPV, CV** | **Cu@PEDOT:PSS** | **Porous Polymer matrix** | **1-100000** | **0.13** | **6** | **10 s** | **Artificial Urine** | **This work** |

Abbreviations: **CV** = Cyclic Voltammetry; **DPV** = Differential Pulse Voltammetry; **SWV** = Square Wave Voltammetry; **EIS** = Electrochemical Impedance Spectroscopy; **PAA** = Polyacrylic Acid; **NPs** = Nanoparticles; **SPCE** = Screen-Printed Carbon Electrode; **rGO** = Reduced Graphene Oxide; **GCE** = Glassy Carbon Electrode; **CNT** = Carbon Nanotube; **ABTS** = 2,2'-azino-bis(3-ethylbenzothiazoline-6-sulphonic acid); **MIP** = Molecularly Imprinted Polymer; **CPE** = Carbon Paste Electrode; **CP** = Carbon Paste; **MWCNT** = Multi-Walled Carbon Nanotube; **Inu** = Inulin; **FA** = Folic Acid; **PDA** = Polydopamine; **NB** = Nile Blue; **SCPE** = Screen-Printed Carbon Paste Electrode; **ITO** = Indium Tin Oxide; **Ni@PANI** = Nickel nanoparticles embedded in Polyaniline; **MCGE** = Modified Carbon Graphite Electrode; **CB** = Carbon Black; **CdS** = Cadmium Sulfide; **PGE** = Pencil Graphite Electrode; **CDs** = Carbon Dots; **PEDOT:PSS** = Poly(3,4-ethylenedioxythiophene):Polystyrene sulfonate.

## 2. Methods

### 2.1 Material and methods

#### 2.1.1 Reagents

Deionized (DI) water with verified resistivity of 18.2 MΩ.cm was obtained using a purification system from Millipore, US. Phosphate buffered saline (PBS) was made from NaCl, KCl, $Na_2HPO_4$ and $KH_2PO_4$ and adjusted to pH 7.4 with HCl, CT concentrations were prepared in PBS prior to the experiments. Copper sulfate pentahydrate ($CuSO_4 \cdot 5H_2O$) solution of 10mM was prepared in PBS, interfering chemicals such as urea, uric acid, NaCl, L-ascorbic acid were obtained from Merck Sigma Aldrich (Poland), components for developing in-house artificial urine media (AUM) are summarized in Supplementary Table S1, monomer 3,4-ethylenedioxythiophene (EDOT), poly(sodium styrene-4-sulfonate) PSS, Benzoyl Peroxide (dibenzoyl peroxide) were also obtained from Merck Sigma Aldrich (Poland).

#### 2.1.2 Sensor platform

For the experimental platform, carbon SPEs (model: 11L) were obtained from DropSens, MetroOhm (Warsaw, Poland). Each SPE consists of one carbon working electrode (diameter ~ 0.4 cm), one carbon counter electrode and one silver/silver chloride (Ag/AgCl) pseudoreference electrode, each connected to individual contact pads through Ag interconnecting lines.

### 2.1.3 Electrode modification

#### 2.1.3.1 PEDOT:PSS synthesis and solution preparation

For the synthesis of PEDOT:PSS, 3.8 ml of PSS (poly(styrene sulfonate)) was dissolved in 50 ml of deionized water. The resulting solution was divided into two equal parts. In one part, 0.6 ml of EDOT was added. In the other part, 3 g of benzoyl peroxide was added. The two solutions were then combined and stirred together for 24 hours inside an incubator with a constant temperature of 37°C. After 24 h, the resulting mixture was removed from the incubator and allowed to cool down to room temperature.

#### 2.1.3.2 PEDOT:PSS functionalization and Cu embedding

For the surface modification of carbon electrodes, 25 μL of as-polymerised PEDOT:PSS was drop casted on the working electrode of the screen-printed electrodes (Metrohm 11L, DropSens) as shown in Supplementary Fig. S1. Post drop casting, the SPE's are oven baked at 75 °C for 24 h. 60 μL of 10 mM $CuSO_4$ solution is pipetted on the sensor region followed by 5 potentiodynamic electrodeposition cycles (cyclic voltammetry) between -1 V to +1 V, at a scan rate of 100 mV/s. After electrodepositon, the sensor was rinsed thoroughly three times with pure PBS solutions to remove any residual $CuSO_4$. This procedure ensured the successful embedding of CuNPs into the conductive PEDOT:PSS matrix resulting in the final fabricated sensor as shown in Supplementary Fig. S2. Prior to analyte testing, the sensors were stored at room temperature under dry conditions. Subsequently, the sensors were connected to a radio-frequency shielded connector attached to a potentiostat to ensure low-noise measurements (Supplementary Fig. S2).

### 2.1.4 Electrochemical characterization

Cyclic voltammograms for different samples were acquired using Bio-Logic SAS (Model : VSP) potentiostat by sweeping the potential from -1 to +1 V vs. Ag/AgCl at a scan rate of 100 mV s$^{-1}$. For each sample, 5 continuous cyclic scans were performed to ensure the data stability. Similarly, differential pulse voltammograms were recorded between –0.5 and +0.5 V vs. Ag/AgCl using 5 mV step size, pulse width ∼ 100 ms and pulse height - 2.5 mV.

### 2.1.5 FESEM and EDX measurements

For the morphological characterization of the samples, examinations were performed under a high vacuum (pressure $10^{-7}$ mbar) with a scanning electron microscope (FEI Nova NanoSEM 450). Images were obtained at a long scan acquisition time (20 μs) of typically 30 seconds per frame for all the samples after carefully selecting the modification region. The FESEM was equipped with EDX detector which is used to perform measurements on individual samples to observe any nature of variation in their chemical composition.

### 2.1.6 XPS measurements

XPS measurements were performed with a Kratos Axis Supra spectrometer (Kratos Analytical Ltd, UK), equipped with a monochromatic Al Kα radiation source (1486.7 eV). All data were collected with an analyzer with pass energy of 80 eV for the survey scan and 20 eV for region spectra, respectively. The instrument work function was calibrated to give a BE of 84.0 ± 0.1 eV for the 4f7/2 line of metallic gold and the spectrometer dispersion was adjusted to give a BE of 932.6 eV for the Cu 2p3/2 line of metallic copper. The effect of sample charging was reduced by a co-axial neutralization system. The occurring shift of the energy scale was corrected by setting the main component of C1s at the literature value of 284.8 eV for adventitious carbon[50].

### 2.1.7 FTIR measurements

FTIR spectra (in middle IR spectral range) were measured with Fourier Transform Infrared Spectrometer Vertex 80 V (Bruker Inc., USA). The spectrometer was equipped with specular reflectance accessory GS19650 (SPECAC, UK). In order to obtain good signal-to-noise ratio, apparatus spectral resolution was maintained to 2 cm$^{-1}$, number of scans performed were 1024 and pressure inside the spectrometer was restricted to < 6hPa. In order to limit the spot size if infrared radiation to the active dimension of electrode, the angle of incidence / reflection was set to 30º.

### 2.1.8 AFM measurements

In this section, experiments were performed by employing MultiMode 8 atomic force microscope equipped with J scanner under the control of NanoScope V controller from Bruker (Bruker Corporation, Santa Barbara, CA) with PeakForce Quantitative Nanomechanical Mapping (PF-QNM) mode was used for samples topography imaging. The RTESPA 300 (Bruker) cantilevers of 40 N m$^{-1}$ spring constant and resonance frequency of 300 kHz were used for AFM imaging in air. The shape of the cantilever was rectangular, with a nominal radius<10 nm. Data analysis was performed using NanoScope Analysis 1.8 from Bruker.

## 3. Results and Discussion

### 3.1 Electrode surface modification and Cu nanoparticle embedding

To develop a CT detection system, we prepared a method whereby CuNPs can be embedded inside the PEDOT:PSS matrix coated on top of carbon working electrodes by employing potentiodynamic electrodeposition technique. Compared to conventional nonenzymatic CT detectable biosensors (**Table 1**), this method enables direct and rapid embedding of CuNPs onto the PEDOT:PSS interstitial sites allowing targeted CT complex formation. Fig. 1 depicts the schematic of the present research work initiated by the process of oxidative polymerisation of EDOT monomer to PEDOT:PSS polymer (**Fig. 1a** and See Methods) which is subsequently drop-casted on top of carbon working electrode as shown in Fig. 1b. After baking the coated devices at 75 ºC for 24 h, we performed an electrochemical sweep from -1 to +1 V vs. Ag/AgCl at a scan rate of 100 mV/s with phosphate buffer saline (PBS) droplet placed on top of the three electrodes. Such voltammogram is then compared with reference carbon-PBS voltammogram (Supplementary Fig. S3) which shows a two-order increase in current values for the PEDOT:PSS modified device. PEDOT:PSS functions as a conductive polymeric matrix that, under applied electrochemical bias, enables the electrochemical deposition of metal nanoparticles (e.g., CuNPs) onto its surface and within its porous structure[49]. This incorporation of nanoparticles substantially increases the electroactive surface area, thereby enhancing the electrode's current response observed in voltammetry.

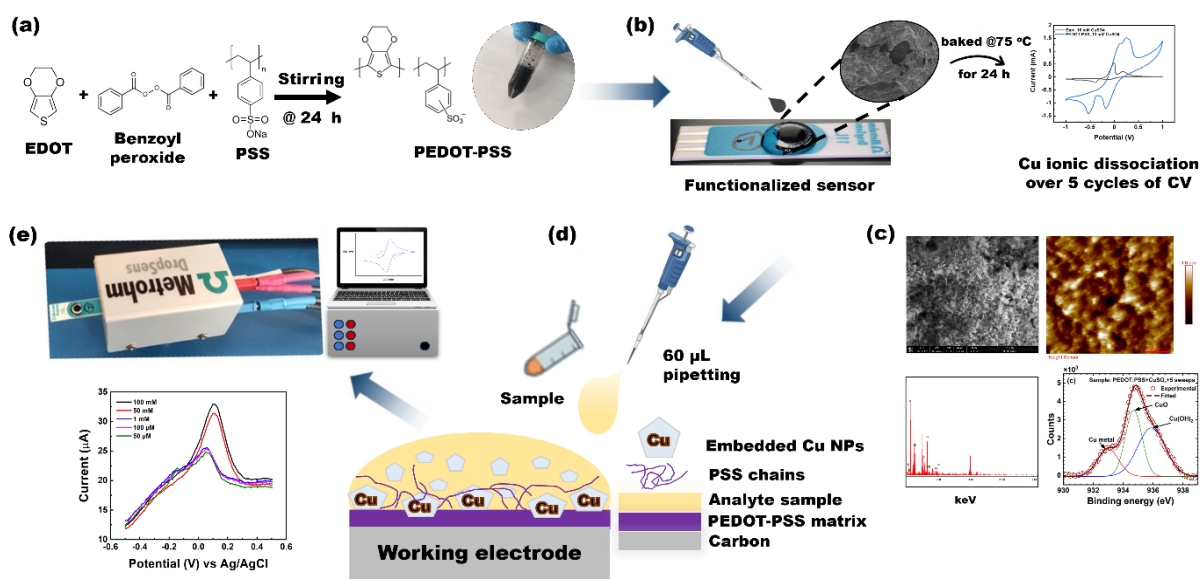

**Fig. 1**: Scheme of the workflow a) Synthesis of PEDOT:PSS, b) Modification of SPE with PEDOT:PSS and Cu embedding via electrochemical potential sweeping, c) characterization of the SPE to confirm embedding of Cu in the conductive polymer matrix, d) Schematic of the sensor region and e) electrochemical measurements and output differential pulse voltammogram for CT detection.

Post PEDOT:PSS modification, the first step is to assess whether Cu can be embedded inside the pores of the polymer matrix. Recent literature survey reports Cu in various phases (nanoparticles, cuprites and oxides) can be incorporated in polymers and form complexes with CT[51,52], however, such processes are quite complicated and expensive. Further, the nature of such complexes is still not clarified yet which results in discrepancies related to real time CT detection for CKD monitoring. We addressed this issue by directly embedding Cu inside the PEDOT:PSS matrix where copper sulfate ($CuSO_4$) mixed with PBS solution is drop casted on top of PEDOT:PSS modified sensor and subjected to 5 electrochemical sweeps (cyclic) from potential (E) -1 to +1 V vs. Ag/AgCl with a scan rate of 100 mV/s. The next step involves removing $CuSO_4$ remnants from the electrode's surface through PBS rinsing followed by the acquisition of CV and DPV data in presence of bare PBS sample. Further, to test the embedding of Cu inside PEDOT:PSS matrix, we performed validatory Field-Emission Scanning Electron Microscopy (FESEM), Energy-Dispersive X-ray Spectroscopy (EDX), X-ray Photoelectron Spectroscopy (XPS) and Atomic Force Microscopy (AFM) experiments as shown in Fig. 1c. Following the fabrication of the Cu embedded PEDOT:PSS sensor, 60 μL samples containing different concentrations of CT are placed on top of the 3-electrode setup as shown in Fig. 1d and their respective CV and DPV data was acquired (Fig. 1e) to assess the variations in output peak current values. From such difference in peak current values for a variety of CT concentrations corresponding to normal (2-20 mM) and chronic (>20 mM) renal failure stages, the final CT concentrations in test samples can be validated and recorded to form a clinical database.

CT is electrochemically inactive as per literature reports suggest[23,52], which also has been validated in this study by performing CV of pure CT in PBS solution placed on top of bare carbon electrodes (Supplementary Fig. S4). However, certain metals can form complexes with CT inside a solution, thereby turning an electrochemically inactive CT to "electrochemically active" as a complex, which can act as an indirect route to detect and monitor CT levels. Following such route, we performed CT detection in artificial urine media by studying the voltammograms of the tripartite complex using a three-electrode sensor chip. **Fig. 2a** shows a comparative CV of a 60 μL droplet of 10 mM $CuSO_4$ mixed with PBS in bare and PEDOT:PSS modified electrodes. In addition to the amplification in base current values, both anodic and cathodic scans depict a substantial increment in the oxidation and reduction peaks of Cu in PEDOT:PSS modified electrode as compared to bare one. The two oxidation peaks at ~ 0.1 V and ~ 0.25 V are attributed to the stepwise oxidation of embedded copper: the first corresponding to $Cu^0 \rightarrow Cu^+$, and the second to $Cu^+ \rightarrow Cu^{2+}$. These species are stabilized within the PEDOT:PSS matrix, likely via electrostatic interactions with sulfonate groups from PSS. It is also observed from **Fig. 2a** that amplification of the second oxidation peak surpassed the first in contrast to the bare. This increment in the $Cu^{2+}$-PEDOT:PSS complex peak suggests an increase in the number of $Cu^{2+}$ ions getting attached to the sulfonate chains of PEDOT:PSS. It is imperative to note that the conductivity enhancement of the PEDOT:PSS matrix is correlated to the softness parameter of the metal ions[53,54]. $Cu^{2+}$ ions have positive softness parameter, hence, it binds strongly with the PSS chains that significantly enhances the conductivity of the PEDOT:PSS film. $Cu^{2+}$ and $SO_4^-$ ions in the PEDOT:PSS matrix screen the charges on PEDOT and PSS, as salts induce charge screening and conformational changes on polyelectrolyte complexes such as PEDOT:PSS[53,54]. The charge screening weakens the coulombic attraction between PEDOT and PSS, followed by the loss of some PSS chains which influences the conductivity enhancement of the PEDOT:PSS matrix. This validates the increment in the second oxidation peak in our PEDOT:PSS modified device. DPV plot of 10 mM $CuSO_4$ in PBS (Supplementary Fig. S5) also shows two distinct peaks representing both $Cu^{1+}$-PEDOT:PSS and $Cu^{2+}$-PEDOT:PSS complexes. Post Cu embedding and PBS rinsing process, the acquired DPV spectra of PBS sample, as shown in **Fig. 2b**, also categorically depicts both the peaks corresponding to such Cu@PEDOT:PSS

complexes, represented by Z I and Z II, respectively. This is attributed to the Cu nanoparticles getting embedded inside the polymer matrix, which would not be washed away even after rinsing the electrode.

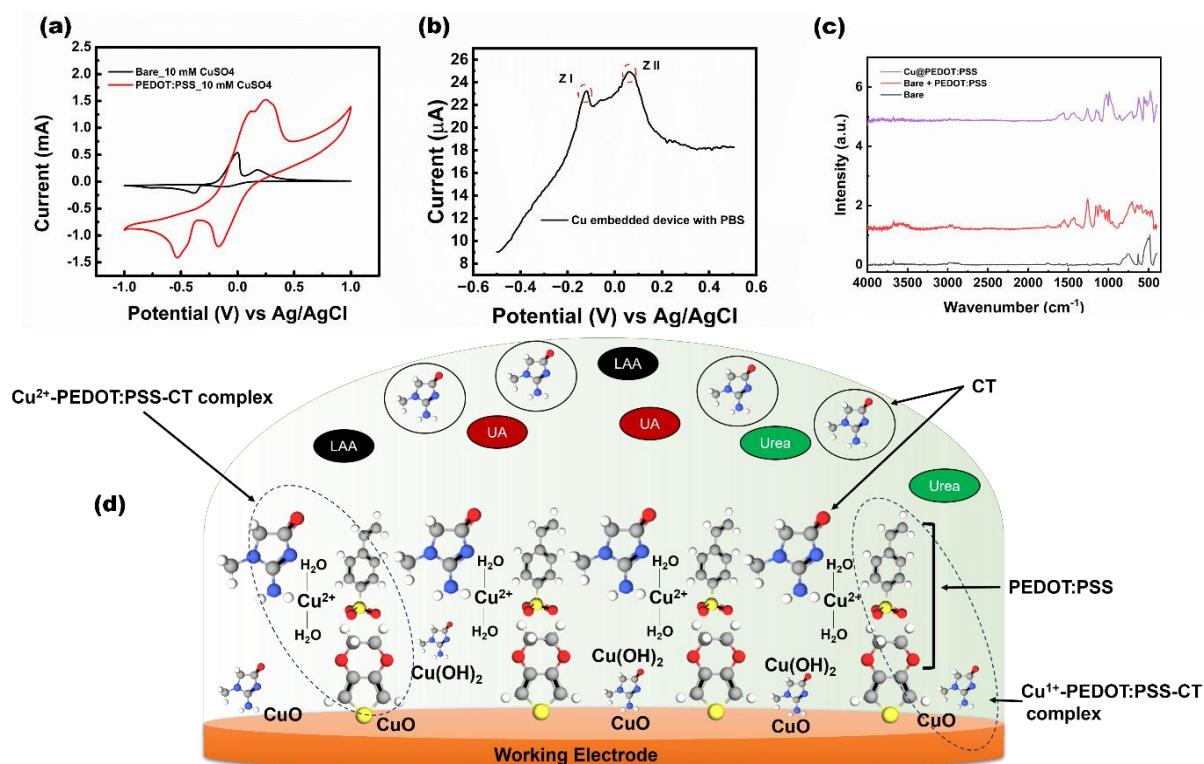

**Fig. 2**: a) Comparative CV of 10 mM CuSO$_4$ on bare and PEDOT:PSS modified SPE. b) DPV voltammogram of a Cu embedded device in PBS c) FTIR spectra of the various stages of biosensor fabrication. d) Schematic of the tripartite complexes forming at the working electrode interface.

**3.2 Characterisation and validation of Cu embedding and understanding of the tripartite complex**

Fourier Transform Infrared Spectroscopy (FTIR) of bare carbon, PEDOT:PSS modified and Cu@PEDOT:PSS devices were performed to assess the presence of any covalent bonds arising due to the formation of such complexes. **Fig.** 2c shows comparative FTIR spectra of such modified layers which depicts the formation of their respective bonds. FTIR spectrum of PEDOT:PSS contains peaks from C-H, S=O and S-C phenyl bonds in sulfonic acid at 1426, 1463, 1153, 1121, and 1019 cm$^{-1}$, and peaks from C=C, C-C, and C-S bonds in the thiophene backbone at 1550, 950, 855, and 709 cm$^{-1}$ respectively [55]. However, post assessing the detailed summary of all the bonds and their corresponding assignments (Supplementary Table S2), no new covalent bonds are observed in the Cu@PEDOT:PSS device and after its interaction with CT (Supplementary Fig. S6). The most intriguing part of this CT detection scheme is to have a comprehensive understanding of the formation of the 'tripartite complex' at the working electrode and its interaction with different interfering complexes present in urine sample. Herein, we focussed on the complexities related to an artificial urine sample droplet placed on top of the working electrode of our fabricated sensor device. Fig. 2d shows the schematic of such droplet consisting of major interfering components like Urea, Uric Acid (UA) and L-Ascorbic Acid (LAA) related to CT detection. Artificial Urine Medium used in the present work is developed in-house (See Supplementary Table S1) following standard protocol[56]. In terms of selective sensing, Cu ions (Cu$^{1+}$ and Cu$^{2+}$) pose an affinity towards binding with CT molecule[57] and, it was interesting to observe how already formed complexes (Cu$^{1+}$-PEDOT:PSS and Cu$^{2+}$-PEDOT:PSS) acted in presence of the CT molecules. We predicted, the most common sources of Cu after CuSO$_4$ dissociation are copper oxide (CuO) which is expected to be found on the surface of the PEDOT:PSS, copper hydroxide (Cu(OH)$_2$) molecules inside the solution and Cu nanoparticles which are embedded inside the polymer matrix crosslinked at the interstitial sites/pores of the conjugated polymer.

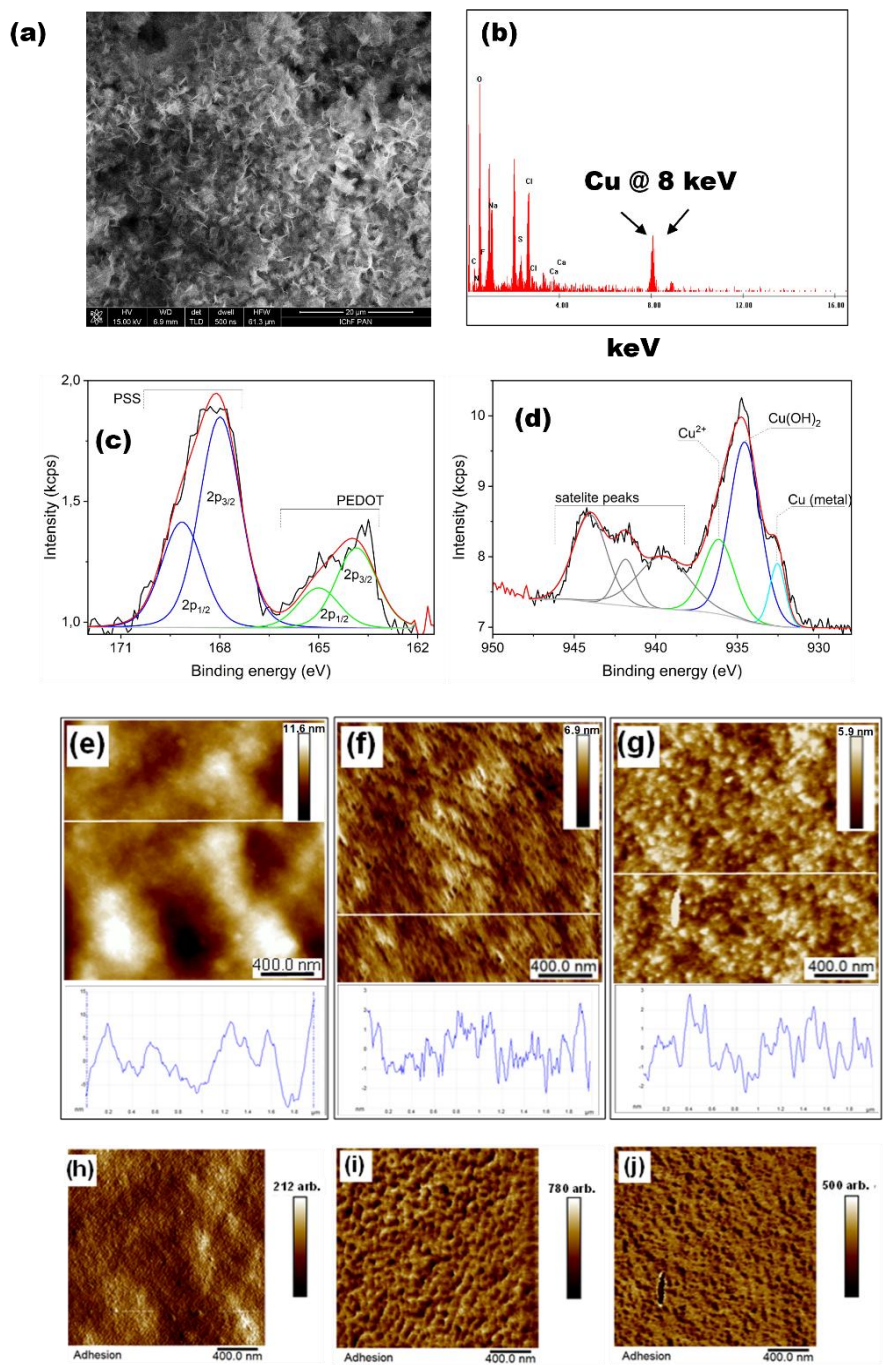

**Fig. 3**: Characterization of the Cu@PEDOT:PSS modified electrode. (a) Field emission scanning electron micrograph of the electrode surface. (b) The EDX analysis shows distinct peaks for Cu at 8 keV. (c) and (d) XPS measurements and analysis of the modified working electrode, (e-g) AFM images and corresponding cross-sectional profiles taken along white lines of (e) PEDOT:PSS, (f) Cu@PEDOT:PSS, and (g) Cu@PEDOT:PSS after interaction with CT, (h-j) Corresponding adhesion maps collecting simultaneously with the AFM images.

Field Emission Scanning Electron Microscopy (FESEM) was performed to understand the morphology of the Cu@PEDOT:PSS complex, as shown in Fig. 3a. Such micrograph depicts a clustered morphology of the conjugated polymer matrix which act as potential sites containing embedded Cu. As compared to the SEM micrographs of bare and PEDOT:PSS (Supplementary Fig. S7 a, b), it is quite evident that such clustered morphology arises due to the PEDOT:PSS matrix being treated with $CuSO_4$. Post CT addition, such clusters re-orient themselves into globular morphology (Supplementary Fig. S7 c) which is attributed to the attachment of CT within the Cu@PEDOT:PSS matrix. Further, Energy Dispersive X-Ray Spectroscopy (EDX) was carried out in-situ under FESEM chamber to verify the presence of Cu, with particular focus on such potential sites. Fig. 3b indicates the presence of metallic Cu corresponding to a sharp peak at 8 keV along with the presence of other elements such as Na, Cl, S, C with their respective sources being PBS, PEDOT:PSS and the carbon electrode itself. The EDX spectra of Cu@PEDOT:PSS is compared with both the bare and PEDOT:PSS modified surfaces (Supplementary Fig. S7 d, e) which depicts no such peak at 8 keV. Fig. 3c–d show the X-ray photoelectron spectroscopy (XPS) results for the Cu-embedded sample. In Fig. 3c, the typical S 2p signatures of PEDOT at a binding energy of 164 eV and of PSS at 168 eV are observed[58]. The presence of Cu was analyzed in the Cu $2p_{3/2}$ region (950–928 eV), as shown in Fig. 3d. Peak fitting of the spectra revealed multiple components. Three of them, at binding energies of 932.8, 934.8, and 935.8 eV, are assigned to metallic copper, $Cu(OH)_2$, and $Cu^{2+}$ species, respectively[59]. It should be noted that the presence of CuO, which exhibits a similar binding energy, cannot be excluded. However, additional analysis of Auger-type transitions (Fig. S8) indicates that the main signal appears at a kinetic energy of approximately 916 eV, which is expected for $Cu(OH)_2$ or $CuSO_4$. In contrast, the characteristic signal for CuO should dominate at 920 eV. A weaker but distinct peak at 922 eV is characteristic of metallic copper, confirming its correct assignment in the Cu $2p_{3/2}$ spectrum. The remaining group of components in the 938–948 eV range corresponds to satellite features that are typically observed for copper in the +2 oxidation state[59]. **Fig.** 3(e-j) further show atomic force microscopy (AFM) images of the carbon electrode modified with PEDOT:PSS polymer (e) before and (f) after Cu embedding while **Fig.** 3g shows the AFM image of Cu@PEDOT:PSS after interaction with CT. Cu trapping in the PEDOT:PSS polymer matrix significantly alters the morphology of this polymer. The morphology of PEDOT:PSS before Cu trapping is a homogeneous porous structure with an average surface roughness (r.m.s) of $R_s$ = 1.35 nm. The embedding of Cu ions into the polymer results in a significant decrease of the roughness value to $R_s$ = 0.76 nm. This is clearly evident in the example cross-sectional profiles shown below each of the AFM images. The resulting Cu@PEDOT:PSS composite exhibits a more compact structure with small, circular pores of ~30 nm in diameter. Further treatment of the Cu@PEDOT:PSS sample with CT caused the polymer morphology to become more granular with a roughness of $R_s$ = 0.78 nm. The morphology changes visible on these AFM images are accompanied by viscoelastic changes in the polymer, resulting in an increase in adhesion (**Fig.** 3h-j), suggesting that the polymer became more viscous to the AFM probe after Cu addition. Thus, such analyses validate successful embedding of Cu inside the polymer matrix, which can be further exploited to detect different CT levels in urine samples.

**3.3 Electrochemical detection of different CT levels in PBS**

We next tested different samples of spiked CT levels in PBS samples employing CV and DPV techniques, on-chip. CV and DPV spectra of blank PBS were measured prior to the spiking of calibrated CT concentrations. Further, CT concentration levels such as 20, 40, 60, 80, 100 mM associated to chronic stages of renal failures are subjected to CV and DPV measurements.

CV spectra of PBS samples spiked from 1 µM to 100 mM CT (Supplementary Fig S9) shows distinct alterations in peak current values, however, it is not very specific towards CT level detection. Interestingly, DPV spectra showed distinct changes in output peak current values on addition of different CT levels in both Z I and Z II, however, the nature of change is observed to be opposite. Fig. 4a depicts individual DPV plots corresponding to different CT levels which indicates both the $Cu^{1+}$-PEDOT:PSS-CT and $Cu^{2+}$-PEDOT:PSS-CT complex peaks for Z I ~ -0.2 V and Z II ~ 0.15 V, respectively. It is also apparent from such figure that, there is a gradual disappearance of the oxidation peak current in Z I as compared to Z II, where the increment is prominent with increase in CT levels. To venture further into the spectra, we deconvoluted both the peaks for Z I and Z II (Figs. 4b and 4c) which shows a more clarified picture of the nature of changes occurring at such potentials with increase of CT levels in PBS.

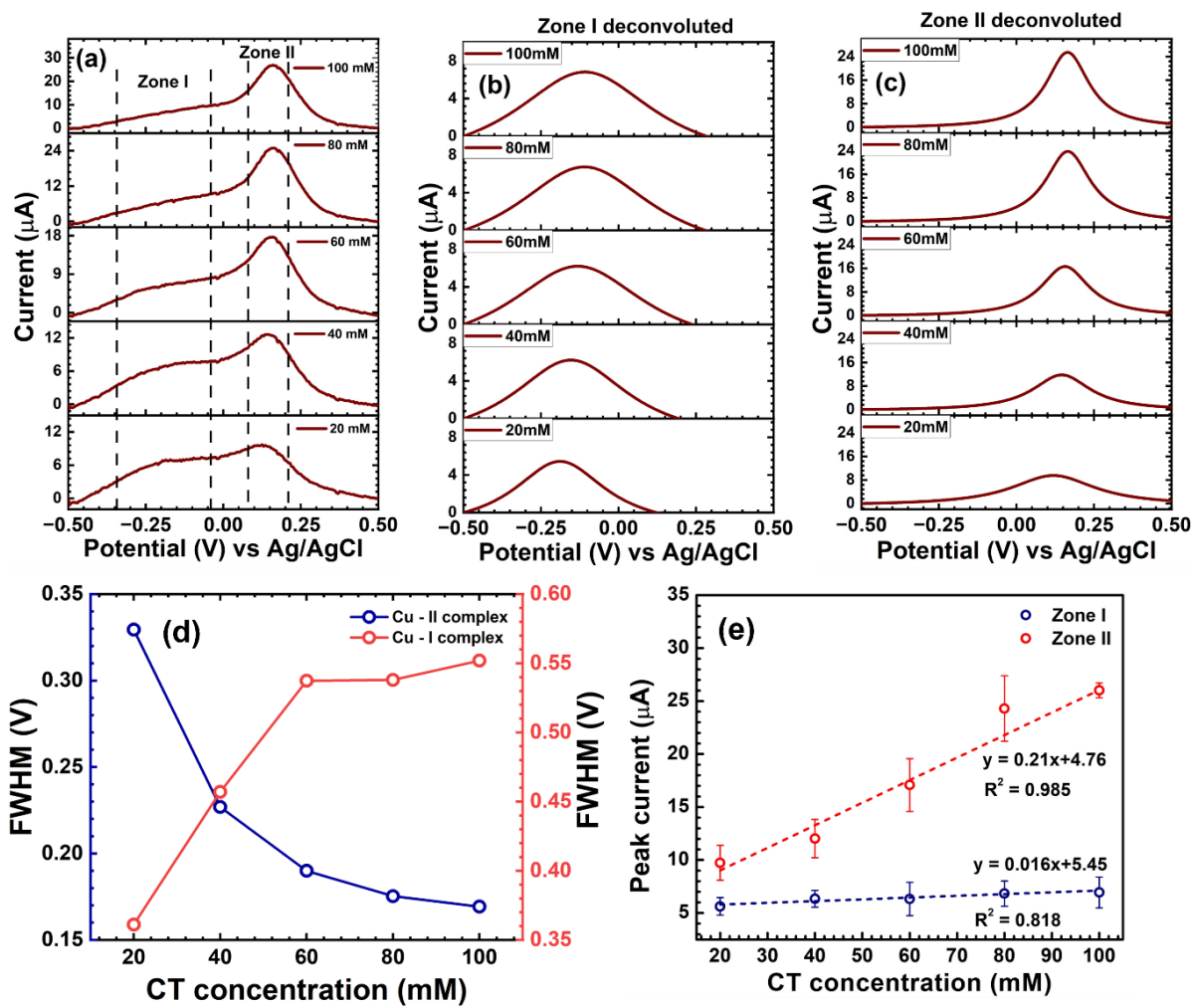

**Fig. 4**: a) DPV spectra showing Zone I and Zone II Cu-PEDOT:PSS-CT oxidation peaks for CT concentrations 20, 40, 60, 80 and 100 mM, b) deconvoluted DPV peaks in Zone I for such concentrations c) deconvoluted DPV peaks in Zone I for such concentrations, d) Full width at half maximum (FWHM) plotted against CT concentration for Zone (I) and Zone (II), e) Calibration curves of zones I and II with error bars and $R^2$ values showing linearity differences.

Deconvoluted Z I shows a slight increment in peak current values ranging from ~5 μA for 20 mM CT to ~7 μA for 100 mM CT in PBS. However, the broadening of the peaks in Fig. 4b from 20 mM to 100 mM is only possible if the number of $Cu^{+1}$ ions are exhausted inside the solution. The steady decrement demonstrates the fact that even if more CT molecules are included inside the solution, the formation of $Cu^{+1}$ ion-based complexes is restricted as their sources CuO and $Cu(OH)_2$ were completely exhausted and therefore, no $Cu^{+1}$-PEDOT:PSS-CT complex formation is occurring henceforth. This phenomenon is further validated by conducting comparative DPV experiments with 1 mM and 10 mM CT in PBS (See Supplementary Fig. S10), where Z I shows a distinct decrement in peak current for increasing CT level, but Z II depicts a complete reverse trend. In Fig. 4c, Z II deconvoluted peaks shows an increment in current values from ~8 μA for 20 mM CT to ~24 μA for 100 mM CT in PBS with increase in CT concentration which validates the continuous formation of $Cu^{2+}$-PEDOT:PSS-CT complexes as the source of $Cu^{2+}$ ions originates from inside the polymer matrix. Also, the narrowing of the peaks suggests the availability of more Cu ions that are still embedded inside the polymer matrix which would take part in complex formation. We further investigated the alteration in peak-width by calculating FWHMs of the peaks for both Z I and Z II to find a suitable sensor readout. It is observed from Fig. 4d that Z I is following a quasi-oscillatory incremental trend with increasing CT concentration whereas, Z II shows an exponential decremental trend, which suggests that FWHM could be a potential to distinguish the respective zones, however, for Z I, it would not be possible to assign FWHM as the sensor readout. On the other hand, Z II makes it suitable for sensing and monitoring of CT levels in a sample which matches our prediction corresponding to Fig. 2d. Apart from FWHM, we tried to evaluate the associated charges corresponding to both Z I and Z II complex formations (See Supplementary Fig. S11) to delve into the formation of different charges with CT concentration increment at every step. However, the results were inconclusive and did not shed enough light on the individual charge calculations. Calibration plots with error bars (**Fig.** 4e) also suggest an upright linearity of Z II ($R^2 \sim 0.98$) as compared to Z I ($R^2 \sim 0.82$) for the chronic CT range, which makes Z II a reliable parameter to detect CT levels in PBS. However, careful observation on Z II anodic peaks suggest a slight shift in their potentials (See Supplementary Fig. S12) which could be due to the oxidation mechanism occurring as a result of one electron transfer from $Cu^{1+}$ to $Cu^{2+}$ during the forward scan.

### 3.4 CT detection in artificial urine media with interference and stability studies

Herein we investigated the effect of different reactive species normally coexisting in human urine such as uric acid, L-ascorbic acid, and urea which could act as potential interfering agents during CT detection and monitoring. Based on the previous experimental results and analyses, we selected Z II as the sole CT detection zone for the electrochemical experiments on artificial urine medium henceforth. DPV voltammograms of AUM in presence of interfering components such as urea, uric acid and L-ascorbic acid showing its representative peak at $\sim 0.14$ V (**Fig.** 5a) which is attributed to the CT level which is already present in the AUM. However, when AUM is spiked with CT concentrations such as 25, 50, 75 and 100 mM, the potential shifts to $\sim 0.16$ V with an increase in peak currents. The increase in peak currents and shifting of the peak potential occurs due to repeated addition of CT along with constant formation of tripartite complexes ($Cu^{2+}$-PEDOT:PSS-CT) due to generation of more $Cu^{2+}$ ions during forward anodic scans.

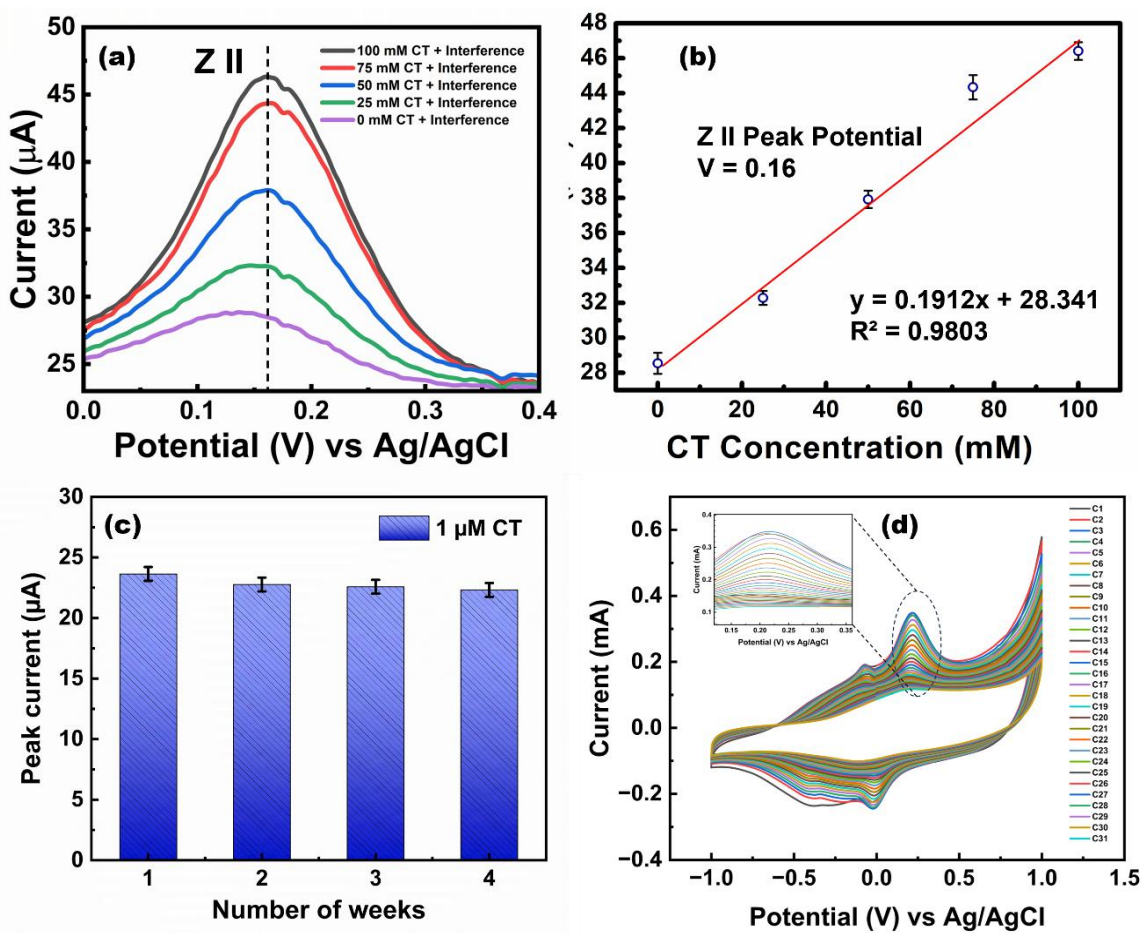

**Fig. 5**: Interference and stability of the fabricated biosensor. a) DPV spectra of AUM along with potential interfering components : Urea, L-ascorbic acid and uric acid, spiked with 25, 50, 75 and 100 mM CT. b) Calibration plot with error bars showing linearity of the measured current values for such CT concentrations at Z II peak potential ~ 0.16 V, c) Electrode stability plot showing peak current variation with time measured for 1 µM CT, d) Cyclic voltammogram of PBS on Cu embedded working electrode measured over 30 cycles, inset shows the decreasing peak current in Z II.

It is imperative to note that, addition of CT in artificial urine media is altering the peak corresponding to CT and no other new peaks are observed in the selected potential range. Further, calibration plot of the current values (**Fig.** 5b) shows linearity ($R^2$ = 0.98) with increase of CT concentration which validates the efficiency of the detection process. The sensitivity and limit of detection (LOD) of the experimental process is obtained to be 0.13 µA.mM cm$^{-2}$ and 6 µM respectively which fits well within the scope of previously reported results. The LOD is calculated using the standard formula LOD = 3σ/β, where, σ is the standard deviation of response at the lowest concentration of CT (1 and 5 µM), and β is the sensitivity factor of the calibration curve. Sensitivity factor represents dy/dx i.e. change in current (I)/ change in CT concentration as extracted from the experimental data for different CT concentrations. Further, to obtain the necessary data repeatability results, one device is subjected to DPV measurement every week for 4 weeks (**Fig.** 5c). The resulting variation in peak current is observed to be 0.53 % after one month which shows excellent electrode stability towards repeated usage. Another study was conducted to virtually peek into the possible quantity of Cu ions embedded inside the conjugated polymer matrix where 30 cycles of redox CV spectra were collected for PBS sample drop-casted on a Cu embedded device (**Fig.** 5d). Such plots reveal a gradual decrement in peak current for both Z I and Z II corresponding to a proportional exhaustion in both $Cu^{1+}$ and $Cu^{2+}$ ions with every passing cycle. This degradation is attributed to the constant formation of both $Cu^{1+}$-PEDOT:PSS and $Cu^{2+}$-

PEDOT:PSS complexes inside the system with increase in cycles. However, the device still seems to work up to 30 cycles showing the corresponding peaks thereby depicting stability of the electrode modification as well as the amount of Cu ions embedded inside the polymer matrix to sustain these electrochemical cycles. The results from this CV can propel further research to determine the exact number of Cu ions embedded in the polymer matrix for further strengthening the reliability factor on the detection and monitoring of CT levels.

## 4. Conclusion

In conclusion, we have successfully fabricated a Cu embedded PEDOT:PSS modified carbon sensor for the detection of CT in artificial urine media employing electrochemical techniques. A simple 5 cycle potentiodynamic electrodeposition using $CuSO_4$ solution on top of PEDOT:PSS modified carbon electrode facilitates the Cu embedding process inside the polymer matrix. Evidence of Cu embedding at the interstitial sites/pores of the conjugated polymer is obtained from EDX, XPS, AFM and electrochemical experimental data. We predicted the formation of two tripartite complexes $Cu^{1+}$-PEDOT:PSS-CT and $Cu^{2+}$-PEDOT:PSS-CT inside the sensor zone which is validated from the electrochemical measured data. PEDOT:PSS modification of the carbon electrode provides a boost in the peak current amplitude corresponding to $Cu^{2+}$-PEDOT:PSS-CT as compared to the bare carbon one. $Cu^{1+}$-PEDOT:PSS-CT is observed to gradually decrease with increasing CT concentration whereas $Cu^{2+}$-PEDOT:PSS-CT increases thereby proving the later to be a perfect candidate for CT detection in real time CT detection and monitoring systems. FWHM study of the oxidation peaks for different CT concentrations show significant potential towards CT detection as compared to the determination of associated ion charge for Z II ($Cu^{2+}$-PEDOT:PSS-CT). Interference experiments show the detection of inherent CT in artificial urine media with no unwanted interference signals from the interfering components. Further spiking the artificial urine media with CT concentrations increase the DPV peak current corresponding to the Cu-PEDOT:PSS CT complexes. Device stability and repeatability studies also confirm that such device can be used for at least a month or more without any degradation in experimental peak current value. Therefore, such novel yet simple concept of a metal embedded conjugated polymer modified electrode can be further used to fabricate biomedical sensor devices for Chronic Kidney Disease monitoring in patients.

**CRediT authorship contribution statement**

**Chirantan Das:** Writing – original draft review & editing, Conceptualization, Methodology, Investigation, Data analysis and curation, Experimentation. **Subhrajit Sikdar:** Writing – original draft review & editing, Methodology, Data Analysis. **Shreyas Vasantham:** Writing – original draft, Experimentation, Data analysis. **Piotr Pięta:** Writing –review & editing, Methodology, Data analysis and curation. **Marcin Strawski:** Writing –review & editing, Data analysis and curation. **Marcin S. Filipiak:** Writing – review & editing, Methodology, Data analysis. **Paweł Borowicz:** Experimentation, Data analysis and curation. **Yurii Promovych:** Experimentation, Data analysis. **Piotr Garstecki:** Writing – review & editing, Data analysis, Supervision, Project administration, Funding acquisition.

**Declaration of competing interest**

The authors declare that they have no known competing financial interests or personal relationships that could have appeared to influence the work reported in this paper.

**Data Availability**

The data that supports the findings of this study are available from the corresponding authors upon request.

**Acknowledgement**

Funding for this research was provided by the National Science Centre, Poland, under the project scheme Maestro 10, 2018/30/A/ST4/00036. CD acknowledges financial support from the Maestro 10 grant. SKV acknowledges support from the Foundation for Polish Science TEAM NET POIR.04.04.00-00-16ED/18-00. The authors are also grateful to the National Science Centre, Poland for the research infrastructure support towards the development of this work.